# A Deterministic Model for Analyzing the Dynamics of Ant System Algorithm and Performance Amelioration through a New Pheromone Deposition Approach

Ayan Acharya[1], Deepyaman Maiti[2], Amit Konar[3]
Department of Electronics and Telecommunication Engineering
Jadavpur University
Kolkata: 700032, India
[1]masterayan@gmail.com, [2]deepyamanmaiti@gmail.com,
[3]konaramit@yahoo.co.in

Janarthanan Ramadoss[4]
Department of Information Technology
Jaya Engineering College
Chennai: 600017, India
[4]srmjana_73@yahoo.com

*Abstract*— Ant Colony Optimization (ACO) is a metaheuristic for solving difficult discrete optimization problems. This paper presents a deterministic model based on differential equation to analyze the dynamics of basic Ant System algorithm. Traditionally, in all Ant System algorithms developed so far, the deposition of pheromone on different parts of the tour of a particular ant is always kept unvarying. This implies that the pheromone concentration remains uniform throughout the entire path of an ant. This article introduces an exponentially increasing pheromone deposition approach by artificial ants to improve the performance of basic Ant System algorithm. The idea here is to introduce an additional attracting force to guide the ants towards destination more easily by constructing an artificial potential field identified by increasing pheromone concentration towards the goal. Apart from carrying out analysis of Ant System dynamics with both traditional and the newly proposed deposition rules, the paper presents an exhaustive set of experiments performed to find out suitable parameter ranges for best performance of Ant System with the proposed deposition approach. Simulations with this empirically obtained parameter set reveal that the proposed deposition rule outperforms the traditional one by a large extent both in terms of solution quality and algorithm convergence. Thus, the contributions of the article can be presented as follows: i) it introduces differential equation and explores a novel method of analyzing the dynamics of ant system algorithms, ii) it initiates an exponentially increasing pheromone deposition approach by artificial ants to improve the performance of algorithm in terms of solution quality and convergence time, iii) exhaustive experimentation performed facilitates the discovery of an algebraic relationship between the parameter set of the algorithm and feature of the problem environment.

*Keywords*— Ant System algorithm, Stability Analysis of Ant System dynamics, Uniform deposition rule, Non-uniform pheromone deposition approach, Solution Quality, Convergence Speed.

## I. INTRODUCTION

**Stigmergy** is a special kind of communication prevalent among some species of ants. While roaming from food sources to the nest and vice versa, ants deposit on the ground a substance called **pheromone**. Ants can detect pheromone and choose, in probability, paths marked by stronger pheromone concentration. Hence, the pheromone trail allows the ants to find their way back to the food source or to the nest. Denebourg *et al.* [1] first studied the pheromone laying and following behavior of ants. **Ant System** (**AS**) ([2]) and **Ant Colony Optimization** (**ACO**) ([3]) owe their inspiration to the works of Denebourg *et al*.

**AS** is the earliest form of ACO algorithm that has been modified by numerous researchers till date to produce many of its variants like **Elitist Ant System** (**EAS**) ([4]), **Max-Min Ant System** (**MMAS**) ([5]), **Rank Based Ant System** (**AS**$_{rank}$) ([6]), **Ant Colony System** (**ACS**) ([7]) etc. Despite the availability of extensive literature on AS algorithms and their applications, very few results are available on its theoretical foundation. The convergence proof of ACO for a graph based ant system by Gutjahr ([8]) also needs special attention. Dorigo and Stutzle in [9] gave a convergence proof for the ACS and MMAS algorithms. Merkle and Middendorf ([10]) studied the behavior of ACO algorithms by analyzing the dynamics of the pheromone model.

This article presents a deterministic model of basic AS dynamics based on differential equation. The analysis helps find out the range of parameters that ensure system's convergence. However, this deterministic model does not violate the stochastic nature of ant system algorithm because ant's trajectory is always chosen by a probability based selection approach. Moreover, the paper presents a novel pheromone deposition approach by artificial ants in which ants increase the amount of pheromone deposition exponentially with time; contrary to the uniform deposition approach that has been used so far in all variants of **AS** algorithms. The underlying philosophy of using this approach is to create a concentration gradient from source to destination in a goal-centric problem which, as if, creates an artificial potential field in the search space and guides ants to find better solution in lesser time. Simulation results reveal that the proposed deposition rule outstrips the traditional uniform deposition approach by a large extent.

The paper is divided into five sections. The deterministic model is presented in section 2 along with the stability analyses of ant system dynamics with both uniform and constant deposition rule. Section 3 presents the experimental results and comparative study of two deposition rules on basic AS algorithm. Finally, conclusions and scope of future work are listed in section 4. We are unable to provide here an introduction to ACO metaheuristics owing to the space constraint.

II. DETERMINISTIC MODEL FOR ANT SYSTEM DYNAMICS

In this section, we provide a simplified analysis of the classical and the extended ant system algorithms with exponential pheromone deposition rule. The objective of this analysis is to determine the parametric conditions for stability of the ant dynamics.

Let, $i$ and $j$ be two successive nodes on the tour of an ant and $\tau_{ij}(t)$ be the pheromone concentration created by the ant at time $t$ and associated with the edge of the graph joining the nodes $i$ and $j$.

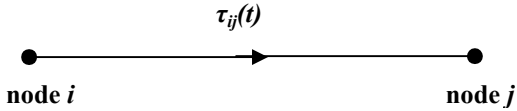

Fig. 1: Defining $\tau_{ij}(t)$

Let $\rho > 0$ be the pheromone evaporation rate, and $\Delta\tau_{ij}^{k}(t)$ be the pheromone deposited by ant $k$ at time t. The basic pheromone updating rule in **AS** is then given by

$$\tau_{ij}(t) = (1-\rho)\tau_{ij}(t-1) + \sum_{k=1}^{m}\Delta\tau_{ij}^{k}(t) \quad (1)$$

From (1), it follows, $\tau_{ij}(t) - \tau_{ij}(t-1) = -\rho\tau_{ij}(t-1) + \sum_{k=1}^{m}\Delta\tau_{ij}^{k}(t)$

$$\Rightarrow \frac{d\tau_{ij}}{dt} = -\rho\tau_{ij} + \sum_{k=1}^{m}\Delta\tau_{ij}^{k}(t) \quad \text{with} \frac{d\tau_{ij}}{dt} = \tau_{ij}(t) - \tau_{ij}(t-1)$$

$$\Rightarrow (D+\rho)\tau_{ij}(t-1) = \sum_{k=1}^{m}\Delta\tau_{ij}^{k}(t),$$

$$\therefore (D+\rho)\tau_{ij}(t) = \sum_{k=1}^{m}\Delta\tau_{ij}^{k}(t+1) \quad (2)$$

where $D \equiv d/dt$ is the differential operator. Evidently, (2) gives the solution for the basic ant system dynamics. Now, to solve (2), we have to solve separately for the complimentary function and the particular integral. We now consider two different forms of $\Delta\tau_{ij}^{k}(t)$ corresponding to both classical AS and modified AS and attempt to determine the closed form solution to $\tau_{ij}(t)$.

• **Evaluation of Complimentary Function**

The complimentary function (CF) of (2) is obtained by setting $\sum_{k=1}^{m}\Delta\tau_{ij}^{k}(t+1)$ to zero. This gives only the transient behavior of the ant system dynamics. Therefore, from (1),

$(D+\rho)\tau_{ij} = 0$, $\Rightarrow D = -\rho$

Thus, the transient behavior of the Ant System is given by:
$\tau_{ij}(t) = Ae^{-\rho t}$ (3)

where, $A$ is a constant which is to be determined from initial condition.

• **Evaluating the Particular Integrals for Different Forms of Pheromone Trail Construction**

The steady-state solution of the ant system dynamics is obtained by computing particular integral (PI) of (2). In this study, we consider two different forms of pheromone deposition: The particular integral (PI) is now given by,

$$\tau_{ij}(t) = \frac{1}{D+\rho}\sum_{k=1}^{m}\Delta\tau_{ij}^{k}(t+1) \quad (4)$$

**Case I:** When $\Delta\tau_{ij}^{k}(t) = C_k$, we obtain from (4)

$$PI = \frac{1}{D+\rho}\sum_{k=1}^{m}C_k = \frac{1}{\rho}\cdot(1+D/\rho)^{-1}\sum_{k=1}^{m}C_k$$

$$= \frac{1}{\rho}(1 - \frac{D}{\rho} + \frac{D^2}{\rho^2} - \ldots)\sum_{k=1}^{m}C_k$$

$$= \frac{1}{\rho}(1)\sum_{k=1}^{m}C_k = \sum_{k=1}^{m}C_k/\rho \quad (5)$$

**Case II:** When $\Delta\tau_{ij}^{k}(t) = C_k(1-e^{-t/T})$, from (4),

$$PI = \frac{1}{D+\rho}\sum_{k=1}^{m}C_k(1-e^{-(t+1)/T})$$

$$= \frac{1}{D+\rho}\sum_{k=1}^{m}C_k - \frac{1}{D+\rho}\sum_{k=1}^{m}C_ke^{-(t+1)/T}$$

$$= \sum_{k=1}^{m}\frac{C_k}{\rho} - \frac{1}{D+\rho}\sum_{k=1}^{m}C_ke^{-(t+1)/T}$$

$$= \frac{\sum_{k=1}^{m}C_k}{\rho} - \frac{\sum_{k=1}^{m}C_ke^{-(t+1)/T}}{(\rho - \frac{1}{T})} \quad (6)$$

For constant deposition rule, the complete solution can be obtained by adding CF and PI from (3) and (5) respectively and is given by,

$\tau_{ij}(t) = Ae^{-\rho t} + \sum_{k=1}^{m}C_k/\rho$.

At t=0, $\tau_{ij}(0) = A + \sum_{k=1}^{m}C_k/\rho$, $\Rightarrow A = \tau_{ij}(0) - \sum_{k=1}^{m}C_k/\rho$

Therefore, the complete solution is,

$\tau_{ij}(t) = \tau_{ij}(0)e^{-\rho t} + \sum_{k=1}^{m}C_k/\rho(1-e^{-\rho t})$ (7)

The above equation clearly demonstrates that for $\rho>0$, the terms having the factor $e^{-\rho t}$ tend to become zero as time t increases. Therefore, the system settles down to its steady state value $\sum_{k=1}^{m} C_k / \rho$ as time progresses. The dynamics, therefore, converges for positive values of $\rho$. Figure (2) provides a plot of $\tau_{ij}(t)$ with varying $\rho$.

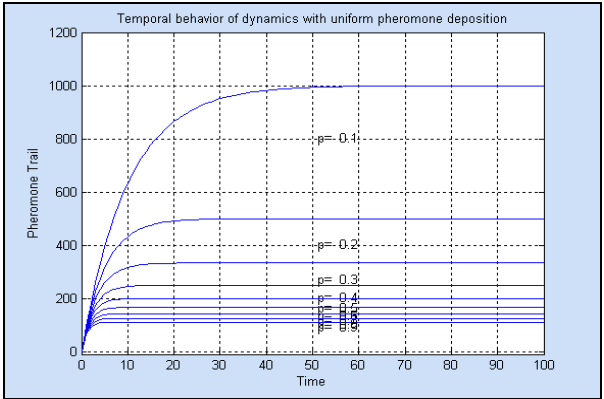

**Figure 2: Plot of $\tau_{ij}(t)$ versus t with varying $\rho$ for constant pheromone deposition**

For exponentially increasing pheromone deposition, the complete solution to the dynamics is as follows,

$$\tau_{ij}(t) = Ae^{-\rho t} + \sum_{k=1}^{m} C_k / \rho - \sum_{k=1}^{m} C_k e^{-(t+1)/T} / (\rho - \frac{1}{T})$$

With initial condition incorporated, the overall solution is given by,

$$\tau_{ij}(t) = \tau_{ij}(0) e^{-\rho t} + \sum_{k=1}^{m} \frac{C_k}{\rho}(1-e^{-\rho t}) + \sum_{k=1}^{m} \frac{C_k e^{-(\rho t + 1/T)}}{(\rho - \frac{1}{T})}(1-e^{-[(1/T)-\rho]t}) \quad (8)$$

Once again, it can be predicted that for positive values of $\rho$, the system converges to its steady state value.

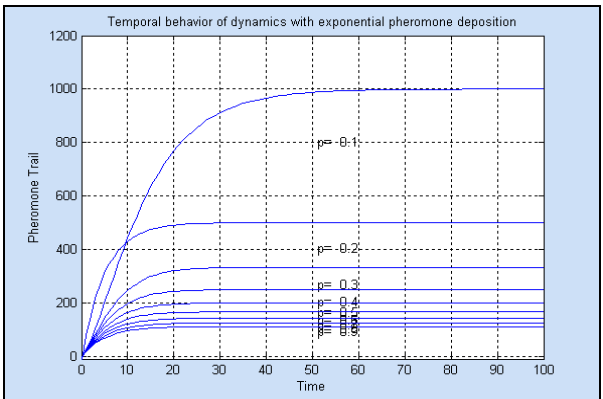

**Figure 3: Plot of $\tau_{ij}(t)$ versus t with varying $\rho$ for exponential pheromone deposition with T=5**

## III. EXPERIMENTAL RESULTS

This section presents the comparative study of two deposition rules. As our problem environment we take a roadmap of connected cities where the shortest route between two given cities is to be determined. We represent the cities as nodes and the paths connecting these cities as edges. Therefore, in effect, the problem environment takes the form of a connected graph $G_C=(C, L)$. $C$ is the set of all nodes and $L$ is the set of all the links which connect two nodes in the graph. Figure (4) shows a sample graph in which the theoretical minimum path, between the source and destination, as found by Dijkastra's algorithm, is shown by red line. Each ant constructs a solution by successively applying the probabilistic choice function ([2]-[6]) which can be described as follows:

$$P_i^k(j) = \begin{cases} (\tau_{ij}^{\alpha}).(\eta_{ij}^{\beta}) / \sum_{k: k \in N_i^k} (\tau_{ik}^{\alpha}).(\eta_{ik}^{\beta}) \text{ if } q<q_0 \\ 1 \text{ if } (\tau_{ij}^{\alpha}).(\eta_{ij}^{\beta}) = \max\{(\tau_{ik}^{\alpha}).(\eta_{ik}^{\beta}): k \in N_i^k\} \text{ with } q>q_0 \\ 0 \text{ if } (\tau_{ij}^{\alpha}).(\eta_{ij}^{\beta}) \neq \max\{(\tau_{ik}^{\alpha}).(\eta_{ik}^{\beta}): k \in N_i^k\} \text{ with } q>q_0 \end{cases}$$

with $P_i^k(j)$ is the probability of selecting node $j$ after node $i$ for ant $k$. $N_i^k$ is the neighborhood of ant $k$ when it is at node $i$. $\eta_{ik}$ is the visibility information generally taken as the inverse of the length of link $(i,k)$. $\tau_{ik}$ is the pheromone concentration associated with the link $(i,k)$. $q_0$ is a pseudo random factor deliberately introduced for path exploration and $\alpha$, $\beta$ are the weights for pheromone concentration and visibility([3],[4]).

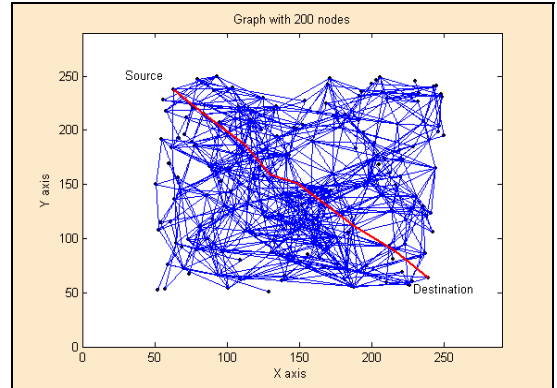

**Figure 4: Graph with 200 nodes**

We divide our simulation strategy in two different levels. In the primary level, we perform experiments over 50 different problem instances to find out a crude range of values of $\alpha$ and $\beta$ for optimum performance of the proposed method. In the secondary level, we vary $\alpha$ and $\beta$ over this crude range and attempt to ascertain an algebraic relation between $\alpha$ or $\beta$ and some feature variables which represent the problem environment. This hierarchical strategy helps in determining exact values of parameter setting for best performance of proposed deposition rule when problem feature set is known in advance.

To save space, we provide here results corresponding to roadmap of 200 cities only. Figures (5) and (6) show that optimum performance with exponential deposition rule is achieved at α=1.0 and β=3.5. In general, it is observed that optimum performance is achieved with α=1.0 and β lying between 3.5 and 4.0. To ensure that the links towards the end of a tour receive the same amount of pheromone as in case of constant deposition rule, we fix *T* at 20% of the average number of links required by ants to construct a tour from source to destination.

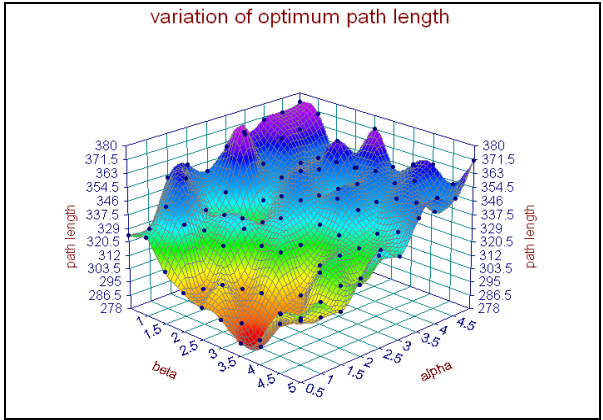

**Figure 5: Variation of optimum path length with α and β**

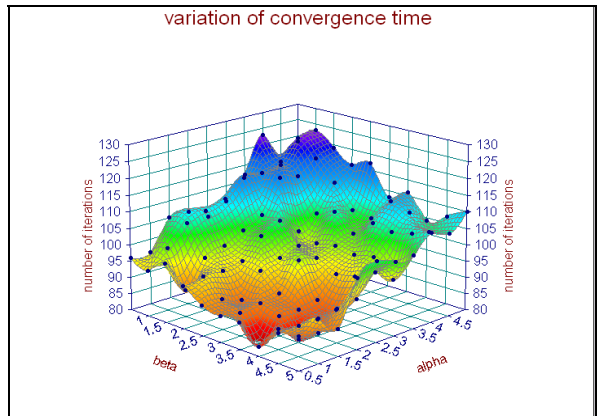

**Figure 6: Variation of convergence time with α and β**

We now vary *α* and *β* over this rough range in steps of 0.1. We identify two parameters to characterize the city distribution. i). **node density (*n*):** signifies the number of cities lying on unit area; ii) **variation coefficient (*σ$_v$*):** which is the standard deviation of the distances of nearest neighbors for all cities divided by average of nearest neighbor distances. In figures (7) and (8), we plot **n**, the number of nodes scattered in an area **300*300** sq. unit along *x* direction and *σ$_v$* along y direction. The functions used in plotting the surfaces along with the co-efficient values are provided underneath each plot.

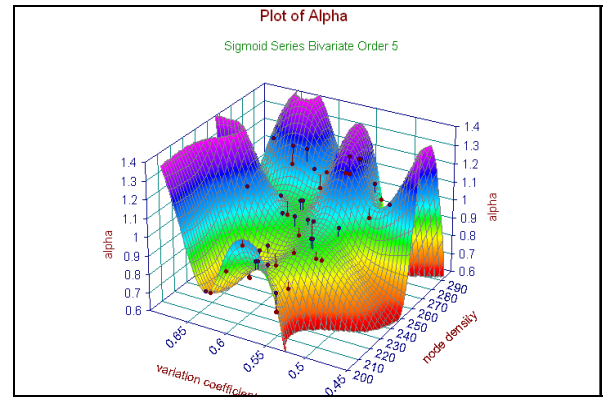

**Figure 7: Plot of *α***
**Function: Sigmoid Series Bivariate Order 5**
*x'*: x scaled -1 to +1; *y'*: y scaled -1 to +1;
$S_{i=2..n}(x')=-1+2/(1+\exp(-(x'+1-(i-1)*(2/n))/0.12))$, $S_1(x')=x'$;
*f(x', y')*=a+

$$\sum_{i=1}^{5} b_i S_i(x') + \sum_{i=1}^{5} c_i S_i(y') + \sum_{i=1}^{4}\sum_{j=1}^{5-i} d_{ij} S_i(x') S_j(y')$$

**Co-efficient Values**
*a*= 0.538, *b$_1$*=-2.167, *b$_2$*=0.903, *b$_3$*=0.479, *b$_4$*=0.215, *b$_5$*=0.410, *c$_1$*=-0.207, *c$_2$*=0.829, *c$_3$*=-0.079, *c$_4$*=0.052, *c$_5$*=0.190, *d$_{11}$*=0.050, *d$_{12}$*=1.319, *d$_{13}$*=0.15, *d$_{14}$*=0.57, *d$_{21}$*=-0.2, *d$_{22}$*=-0.63, *d$_{23}$*=-0.09, *d$_{31}$*=0.027, *d$_{32}$*=-0.18, *d$_{41}$*=-1.022.

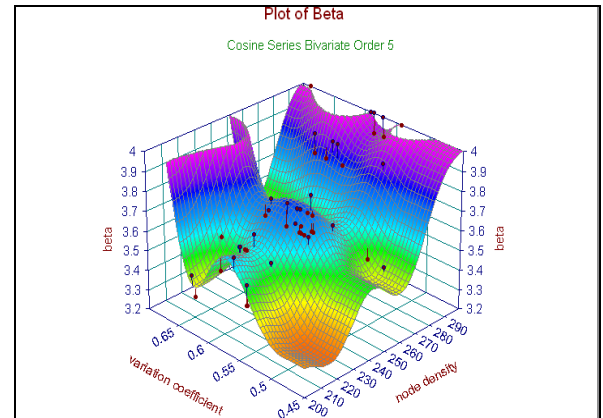

**Figure 8: Plot of *β***
**Function: Cosine Series Bivariate Order 5**
*x"*: x scaled 0 to π; *y"*: y scaled 0 to π;
*f(x", y")*=

$$a+\sum_{i=1}^{5} b_i \cos(ix'') + \sum_{i=1}^{5} c_i \cos(iy'') + \sum_{i=1}^{4}\sum_{j=1}^{5-i} d_{ij} \cos(ix'')\cos(jy'')$$

**Co-efficient Values**
*a*= 3.76, *b$_1$*=-0.06, *b$_2$*=0.07, *b$_3$*=-0.17, *b$_4$*=0.023, *b$_5$*=0.05, *c$_1$*=-0.17, *c$_2$*=0.07, *c$_3$*=-0.11, *c$_4$*=0.03, *c$_5$*=-0.02, *d$_{11}$*=-0.24, *d$_{12}$*=0.122, *d$_{13}$*=-0.159, *d$_{14}$*=0.080, *d$_{21}$*=-0.026, *d$_{22}$*=0.008, *d$_{23}$*=0.002, *d$_{31}$*=0.101, *d$_{32}$*=-0.041, *d$_{41}$*=0.011.

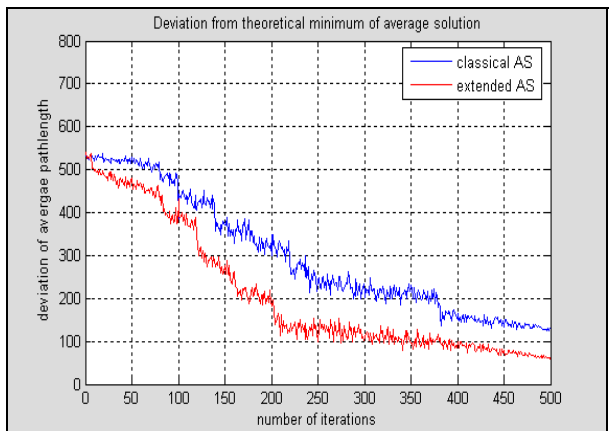

**Figure 9: Comparison of two deposition rules with AS algorithm on a roadmap of 250 cities**

Figure (9) shows a comparative study of deviation of average solution from theoretical minimum of algorithms employing two different deposition rules on a roadmap with 250 cities. As suggested in [3], we take α=1.0, β=2.0 for simulating algorithm with uniform deposition rule. For simulation with exponential deposition, we take help of above equations. For the concerned problem environment, we choose α=1.0, β=3.8. The plots suggest that the average solution is significantly better in case of exponential deposition approach. Moreover, the algorithms with proposed deposition rule exhibit faster convergence.

## IV. CONCLUSIONS AND SCOPE OF FUTURE WORK

The deterministic framework along with the pheromone deposition approach presented in this paper are entirely novel. The analysis of suitable ranges of α and β depending on problem environment has huge practical significance. Our future work will focus on analyzing the dynamics with other variants of AS algorithms and comparative study of two competitive deposition rules on those models.